\documentclass{article}
\usepackage{smc}
\usepackage{times}
\usepackage{ifpdf}
\usepackage[english]{babel}
\usepackage{cite}
\usepackage{amsmath}
\usepackage{algorithm}
\usepackage{algpseudocode}
\usepackage{xpatch}

\makeatletter
\xpatchcmd{\algorithmic}
  {\ALG@tlm\z@}{\leftmargin\z@\ALG@tlm\z@}
  {}{}
\makeatother

\def\papertitle{Generating symbolic music using diffusion models}
\def\firstauthor{Lilac Atassi}

\newif\ifpdf
\ifx\pdfoutput\relax
\else
   \ifcase\pdfoutput
      \pdffalse
   \else
      \pdftrue
\fi

\ifpdf 
  \usepackage[pdftex,
    pdftitle={\papertitle},
    pdfauthor={\firstauthor, 
    },
    bookmarksnumbered, 
    pdfstartview=XYZ 
   ]{hyperref}

  \usepackage[pdftex]{graphicx}
  \graphicspath{{./figures/}}
  \DeclareGraphicsExtensions{.pdf,.jpeg,.png}

  \usepackage[figure,table]{hypcap}

\else 
  \usepackage[dvips,
    bookmarksnumbered, 
    pdfstartview=XYZ 
  ]{hyperref}  

  \usepackage[dvips]{epsfig,graphicx}
  \graphicspath{{./figures/}}
  \DeclareGraphicsExtensions{.eps}

  \usepackage[figure,table]{hypcap}
\fi

\hypersetup{
    colorlinks,%
    citecolor=black,%
    filecolor=black,%
    linkcolor=black,%
    urlcolor=black
}

\title{\papertitle}

\oneauthor
  {\firstauthor} {University of California, San Diego \\ %
    {\tt \href{mailto:latassi@ucsd.edu}{latassi@ucsd.edu}}}

\begin{document}

\emergencystretch 1em

\capstartfalse
\maketitle
\capstarttrue
\begin{abstract}
Denoising Diffusion Probabilistic models have emerged as simple yet very powerful generative models. Unlike other generative models, diffusion models do not suffer from mode collapse or require a discriminator to generate high-quality samples. In this paper, a diffusion model that uses a binomial prior distribution to generate piano rolls is proposed. The paper also proposes an efficient method to train the model and generate samples. The generated music has coherence at time scales up to the length of the training piano roll segments. The paper demonstrates how this model is conditioned on the input and can be used to harmonize a given melody, complete an incomplete piano roll, or generate a variation of a given piece. The code is publicly shared to encourage the use and development of the method by the community.
\end{abstract}

\section{Introduction}\label{sec:introduction}

One of the advantages of generating symbolic music, compared to audio, by machine learning (ML) is that manipulating the generated material is feasible using conventional composition tools. This possibility allows a composer or a musician to collaborate more easily with the ML model. Additionally, using specifically binary piano rolls, i.e., without dynamics, requires less computational resources for training ML models compared to training ML models with spectrograms or even piano rolls with dynamics. The shorter training and sampling time of lightweight models allow for more rapid experimentation. That is the reason that this paper discusses methods that can be applied to binary piano rolls. Nonetheless, most of the presented methods can be generalized to models for other forms of piano rolls, e.g., with dynamics.

The approaches to generative models can be divided into two broad categories. In the first category of approaches, the data distribution is estimated directly. For example, fitting a normal distribution on a set of samples falls under this category. However, this approach does not scale to high-dimensional and complex distributions. Meaning, to increase the accuracy of the estimation of the real data points distribution, an impractical amount of computation is required.

The second approach does not estimate the data distribution directly. Instead, this approach estimates a function that transforms a sample from a prior distribution to the data distribution. The prior distribution is chosen to be simple, that is its probability distribution function has no more than few parameters. Therefore, sampling the prior distribution is straightforward. Generative Adversarial Networks (GANs)~\cite{smc3}, Variational Autoencoders (VAEs)~\cite{smc2}, and Denoising Diffusion Probabilistic (DDP) models~\cite{smc4} or, for short, diffusion models fall under this category. In particular, diffusion models transform a sample from the prior distribution to the data distribution in several gradual steps. 

Multiple methods based on GANs~\cite{smc5} and VAEs~\cite{smc6} have been proposed for music generation in the literature. Mittal et al. in~\cite{smc7} propose using a VAE to encode 32 two-bar segments. The diffusion model in the latent space of the VAE generates new samples. A similar model has also been explored for images~\cite{smc8}. One reason for using a VAE is that the discrete music data is transformed into a continuous latent space where using a diffusion model with a normal prior distribution would be feasible. The proposed method, in this paper, simply generates piano rolls using a binomial prior without the need to transform the data into a continuous space. One advantage of this direct approach is that it is possible to use the model to harmonize a given melody. Or more generally, the model can infill a given piano roll that has some rows and columns masked. The method in~\cite{smc7} can only infill the masked columns. The second advantage of this approach is that the quality of the generated piano rolls is not limited by the decoder of the VAE.

The following sections briefly review the forward and reverse process of the diffusion models, explain the proposed methods for the forward process to generate noisy binary piano rolls, and discuss the proposed sampling algorithm. The experiments and experimental results are presented at the end. The code used in the experiments is available at  \url{https://github.com/lilac-code/music-diffusion}.

\section{Diffusion models}

Diffusion models  \cite{smc4} apply a diffusion transition kernel $T_\pi$,
\begin{align}
    q\left(x_{t}|x_{t-1}\right) &= T_\pi\left(x_{t} | x_{t-1}; \beta_{t}\right) \label{eq:kernel_shorthand},
\end{align}
repeatedly to the input data $x_{0}$ to transform the data distribution to the prior distribution,
\begin{align}
    q\left(x_{0:T}\right) &= q\left(x_{0}\right) \prod_{t=1}^{T} q\left(x_{t} | x_{t-1}\right) \label{eq:diffuse_forward}.
\end{align}
$\beta_t$ is the diffusion rate and $T$ is the number of diffusion steps. The kernel for a binomial prior is 
\begin{align}
B(x_{t}; x_{t-1} (1-\beta_t) + 0.5\beta_t), \label{eq:binomial_kernel}
\end{align}
which can be used for binary vectors such as binary piano rolls. The data generated by the forward process is used to train a neural network that is used in the reverse process, also known as the sampling process. By training the model on pairs of $(x_{t}, x_{0})$, proposed in \cite{smc9}, simply L2 norm can be used as the training loss function. Then, in the sampling process, the model alternates between predicting the noiseless sample $x_{0}$ and adding noise corresponding to $t-1$.

\section{Efficient forward process}
The authors in \cite{smc9} also propose a new equation for the Gaussian kernel in the forward process that depends only on the input $x_0$. This has two benefits. First, to compute the output at time step $t$ there is no need to compute the output at the time steps before $t$. This way, the output at random $t$s can be computed and used to train the neural network for the reverse process. That means with the kernel from the original paper \cite{smc4}, the whole set of noisy training data should be generated and stored in memory. This is needed to shuffle the order of noisy training data in mini-batches, which is required for training neural networks using mini-batch gradient descent. With this updated kernel, it is possible to first pick a random order for the noisy piano rolls and then generate them on the fly without generating the preceding noisy piano rolls. Therefore, before training the network, there is no need to generate all the noisy piano rolls, which is the case in \cite{smc4}. As a result, the amount of memory required is just to hold a single mini-batch and not the whole training set for the neural network. Second, this simplification of the kernel equation, makes it possible to train the neural network to predict the noise between two steps in the reverse process, to predict the mean of the Gaussian kernel, or to predict $x_0$, as discussed in \cite{smc9}. The normal kernel with dependency on $x_{0}$ is presented in \cite{smc9},
\begin{align}
    \alpha_t &= 1 - \beta_t\\
    \bar{\alpha}_t &= \prod_{s=1}^t \alpha_s\\
    q(x_t | x_{0}) &=N(x_t; \sqrt{\bar{\alpha}_t}x_0, (1-\bar{\alpha}_t)I)
\end{align}

To derive a similar binomial kernel with dependency only on $x_{0}$, in Equation~\ref{eq:binomial_kernel}, the success probability can be substituted with the probability from the previous step at $t-1$,
\begin{align}
q(x_t | x_{t-2}) =B\left(x_{t}; x_{t-2} \alpha_t \alpha_{t-1} + (1-\alpha_t \alpha_{t-1}) 0.5\right). 
\end{align}
By repeating the success probability substitution from the prior step until $t=0$, the binomial kernel that depends only on the input $x_0$ is proved to be 
\begin{equation}
    q(x_t | x_{0}) = B(x_t; \bar{\alpha}_t x_0 + (1-\bar{\alpha}_t) 0.5). \label{eq:binomial_kernel_jump}
\end{equation}
Therefore, it is possible to just change the diffusion rate $\beta_t$ schedule in Equation~\ref{eq:binomial_kernel} to arrive at Equation~\ref{eq:binomial_kernel_jump}. In the binomial kernel, $0.5$ is the success probability of the prior binomial distribution. In an application, the average ratio of ones or successes to the dimensionality of $x_0$ replaces $0.5$.

\section{Sampling algorithm}

\begin{algorithm}
\caption{Generating new samples}\label{alg:sample}
\begin{algorithmic}
\State \textbf{Input:} A piano roll sampled from a binomial distribution $x_T$
\For {$t=T,T-1,\ldots,1$}
    \State $\hat{x}_0 = \mathrm{UNet}(x_t)$
    \State $\delta = x_T \oplus \hat{x}_0 $
    \State $\mathrm{mask} \sim B(\delta \beta_t) $
    \State ${x}_{t-1} = \hat{x}_0 \odot (1-$mask$) + x_T \odot \mathrm{mask}$
\EndFor
\end{algorithmic}
\end{algorithm}

\begin{figure}[t]
\centering
\includegraphics[width=1.0\columnwidth]{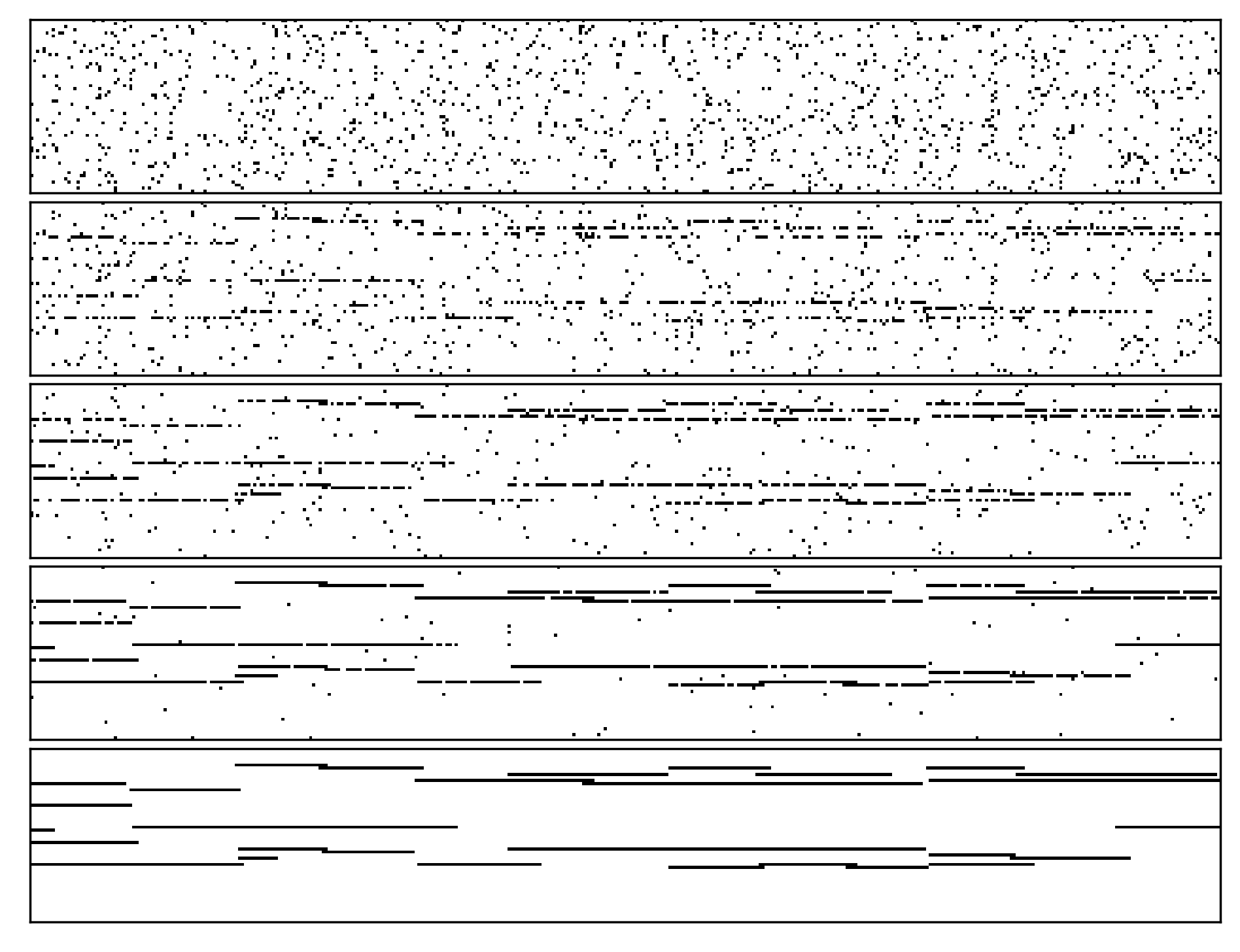}
\caption{Visualizing the the sampled piano roll along the sampling path, $x_t$ in Algorithm~\ref{alg:sample} for $t=100,75,50,25,1$, from top to bottom. It is visually evident the original binomial noise is reduced in each step.\label{fig:sampling_same_noise}}
\end{figure}

The proposed sampling method in this section is inspired by the method in \cite{smc10}. It is shown in \cite{smc10} that when using a neural network that cannot perfectly denoise a given sample, the usual sampling method used with diffusion models can be unstable. As a result, the generated samples are of poor quality. With an improved sampling method, the quality of the generated sample can consistently improve.

In the simple or inefficient sampling method a sample $x_T$ from the binomial distribution is passed to the neural network to estimate the noiseless sample $\hat{x}_0$. Then using the binomial kernel, some noise is added to $\hat{x}_0$ to get $x_{T-1}$. This process of alternating between adding noise and denoising is repeated. However, this method is inefficient as the noise added by the kernel in each step is independent of the noise in the previous step. To see this effect, using the kernel twice with the same $\hat{x}_0$, yields two points that could have a large distance from each other, particularly at $t$s close to $T$. Therefore, after some noise is added to $\hat{x}_0$, $x_{t-1}$ might end up at a point far from $x_{t}$. Consequently, the neural network maps $x_{t-1}$ and $x_{t}$ to two piano rolls that differ vastly. As a result, the computation in those first few steps of the sampling algorithm would be wasted, as each time the target of the mapping by the network is moving around in the space. 

An improved sampling method in each step would partially add back the noise from $x_T$ to $\hat{x}_0$. The proposed method is presented in Algorithm~\ref{alg:sample}. The neural network, in this case a UNet (explained in the next section), is used to denoise the sample. The difference $\delta$ between the denoised sample $\hat{x}_0$ and noise sample $x_T$ is computed using the exclusive or operator. A gradually shrinking subset of $\delta$ in each step is used as a mask that selects what elements of the piano roll should be picked from the noise sample $x_T$. The remaining elements in the piano roll are taken from $\hat{x}_0$. Figure~\ref{fig:sampling_same_noise} shows a piano roll $x_t$ at steps $t=100,75,50,25$ and $1$ generated by the improved sampling method. As the sampling method progresses, it is possible to see in the samples that a subset of the noise remains in the piano roll. As parts of the reconstructed piano roll are preserved in each step, no computation is wasted, and the generated sample quality is improved.

\begin{figure*}[ht]
\centering
\includegraphics[width=0.99\textwidth]{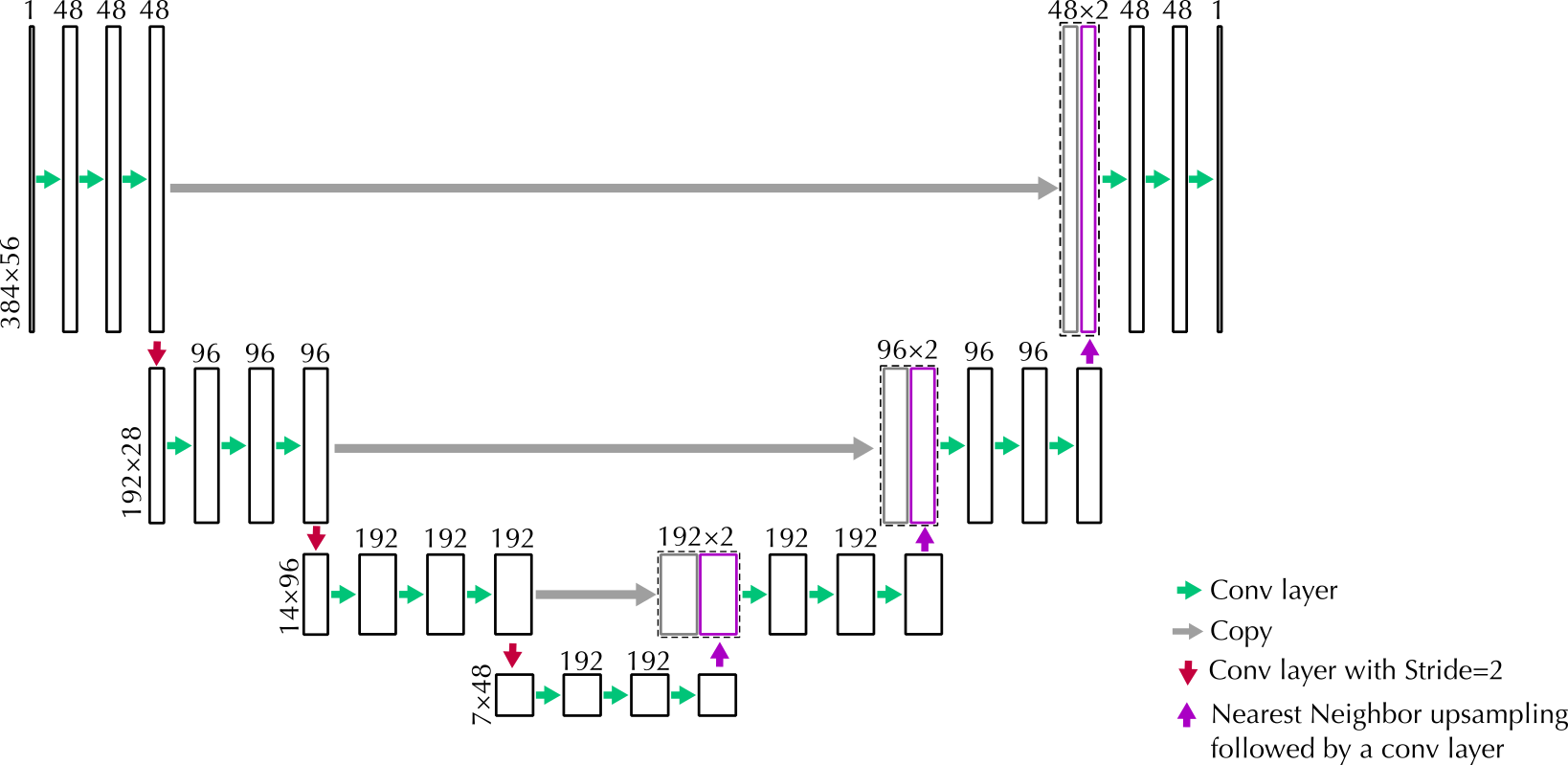}
\caption{This figure depicts the UNet architecture used in the experiments. The input piano roll is a matrix of size $56\times384$. Each green arrow show a conv layer. The rectangles depict the feature maps, with the number above each rectangle showing the depth of the feature-map tensor which is also the number of conv filters in the corresponding layer. The red arrows shows the conv layers with stride of two that reduce the size (rows and columns) of the feature maps by 2. The purple arrows show the upsampling step using the nearest neighbor interpolation, which doubles the size of the feature maps. The grey arrows depict copying a feature map from the left side of the network to the right side. On the right side, the copied feature maps are stacked on the upsampled feature maps, resulting in a feature map with double the depth of the feature-map tensor below. The last conv layer, top right, contains a single convolution filter, as a result, the output is a single piano roll. The number of rows and columns of the tensor in each row or level of UNet remain the same due to padding the input when applying convolution filters.\label{fig:lunet}}
\end{figure*}

\section{Neural network architecture}

The UNet architecture was introduced in 2015 for image segmentation \cite{smc11}. It resembles the typical architecture of autoencoders, called often the hourglass architecture, that subsamples the feature maps in multiple steps in the first half the network and then in the second half the feature maps are upsampled to finally have the identical size of the input. The difference between UNet and hourglass is that UNet has horizontal connections that connect the feature maps at the same scale from the first half of the network to the second half. The UNet architecture used in this paper is illustrated in Figure~\ref{fig:lunet}.

The second main operation in UNet is downampling the feature maps. Downsampling is done by using conv filters with stride of two, skipping every other element of the input tensor horizontally and vertically. Using a conv layer with a stride of two to downsample feature maps has an advantage over just dropping or decimating every other column and row. The learnable weights of the conv filters have an opportunity to generate feature maps that preserve information while reducing the size of feature maps.

The third main operation in UNet is upsampling the feature maps. On the right side of the network, the feature maps are upsampled from higher scales to be combined with the feature maps at lower scales. The upsampling process simply duplicates the values in the input tensor. Upsampling brings the information from higher scales to lower scales, which is used then to guide the finer details.

\section{Training the model}
To train the diffusion model, I used the Maestro dataset that contains MIDI and audio files of recorded performances from piano performance competitions. In the dataset, there are just over 2,000 MIDI files, and the title and composer name for each one are in a text file. Many of the compositions have several recorded performances as MIDI files. I went through the list and removed the duplicated performances to keep only one MIDI file per composition. At this point, about 500 MIDI files were left out of 2,000. Then, using a Python script, I found that there are 78 pieces that have pitches only between 33(A1) and 88 (E6) MIDI notes, which is a range of 56 MIDI notes.

The 78 MIDI files were converted to binary piano-roll matrices with the ones in the matrix indicating what pitch is present at what time tick. The resolution used in the conversion was one quarter note in the MIDI file converted to 24 ticks in the piano roll, which is half the default number of ticks for a quarter note in the MIDI format. Each piano roll was then divided into non-overlapping segments of 16 beats or 384 (16×24) ticks. This process at the end yielded 2,044 piano-roll segments that were used to train the model. The slow training process is the main limiting factor for the training data set size. Improving the training process speed and training on a larger data set remains for future work.

The number of diffusion steps, $T$, used in the experiments was set to 100. The UNet neural network was trained over 50 epochs. In each epoch, each of the 2,044 piano-roll segments was used to generate 100 noisy piano rolls that vary from no noise to all the way being a sample from the binomial distribution. Therefore, in each epoch, there were 204,400 training samples. Training the UNet network on the 50 epochs took
about 48 hours on two NVidia A6000 GPUs.

\begin{figure}[ht]
\centering
\includegraphics[width=1.0\columnwidth]{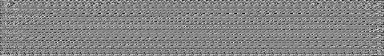}
\includegraphics[width=1.0\columnwidth]{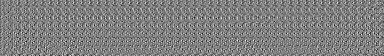}
\includegraphics[width=1.0\columnwidth]{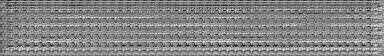}
\caption{Using AM, the input that maximizes the activation of three convolution filters at the lowest resolution of the UNet trained on piano rolls are visualized. The visualized piano rolls reveal the filters at this level, after three downsampling steps, have a high-resolution view of the input piano roll without losing information along the pitch or time axes.}
\label{fig:max_act_1}
\end{figure}

\section{What the network sees}

One question regarding the choice of network architecture is whether the UNet network loses too much information along the pitch axis of the piano rolls each time the input is downsampled. Given that the UNet architecture is designed for processing images and not piano rolls, it is possible that some aspects of the architecture are not well-suited for processing a piano roll with time along one axis and pitch along another, unlike an image that has identical units for the horizontal and vertical axes.

A method that to some degree reveals what each convolution filter in a network is looking at is Activation Maximization (AM) \cite{smc12}. The vanilla AM algorithm takes a random input. In my experiments, this is a piano roll sampled from a binomial distribution. The neural network processes the input.
The gradient of a particular convolution filter’s average output with respect to the input is computed. Then, with gradient ascent, the input is updated. This increases the average of output feature map of the convolution filter. This process is then repeated a fixed number of times.

In my experiments, I found that after about 200 iterations, the updated input does not change much. Applying activation maximization to three convolution filters of the last convolution layer at the bottom of UNet that is trained in my experiments, reveals the convolution filters have a high-resolution view of the input piano roll, as shown in Figure~\ref{fig:max_act_1}. Therefore, the three downsampling steps on the left side of UNet are not losing information along either pitch or time axis. If that was the case, and the convolution filters would have a low-resolution view of the piano roll, the output of AM would have multiple elements along the time or pitch axis with the same color in the repeated patterns. This confirms using convolution filters with stride of two to downsample the feature maps does not loose information as one expects from a simple decimation-based down-sampling method.

\begin{figure}[h]
\centering
\href{https://youtu.be/wu_kfcpzAPI}{\includegraphics[width=0.99\columnwidth]{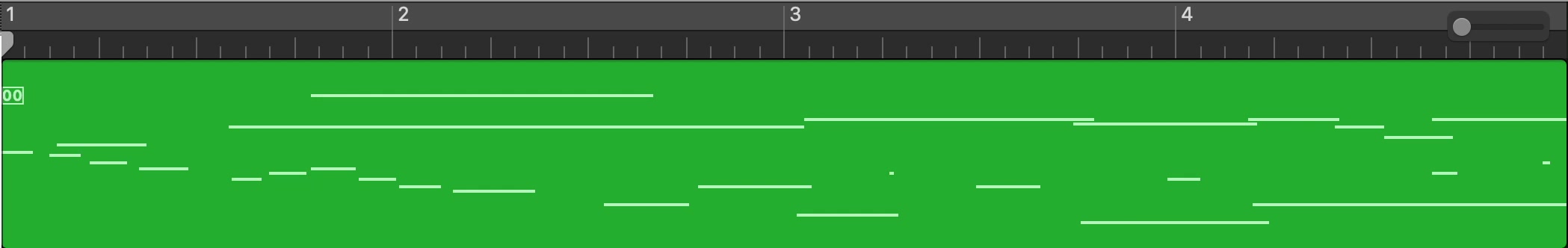}}
\href{https://youtu.be/r3Rn91t3g5A}{\includegraphics[width=0.99\columnwidth]{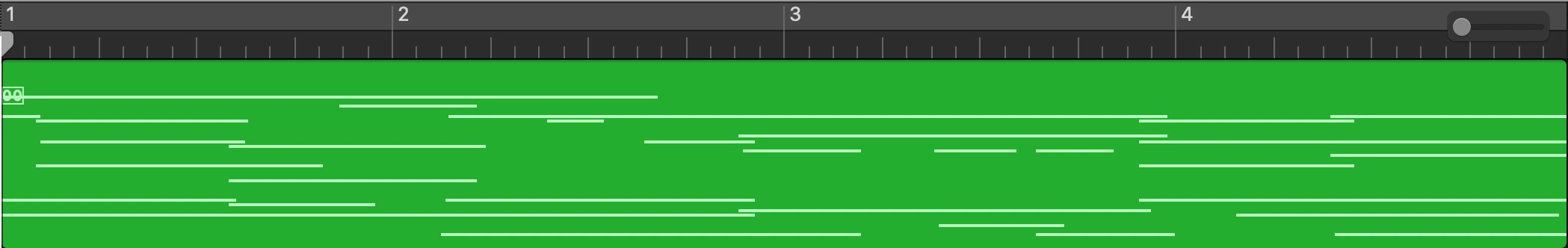}}
\href{https://youtu.be/KEAtbzAScYY}{\includegraphics[width=0.99\columnwidth]{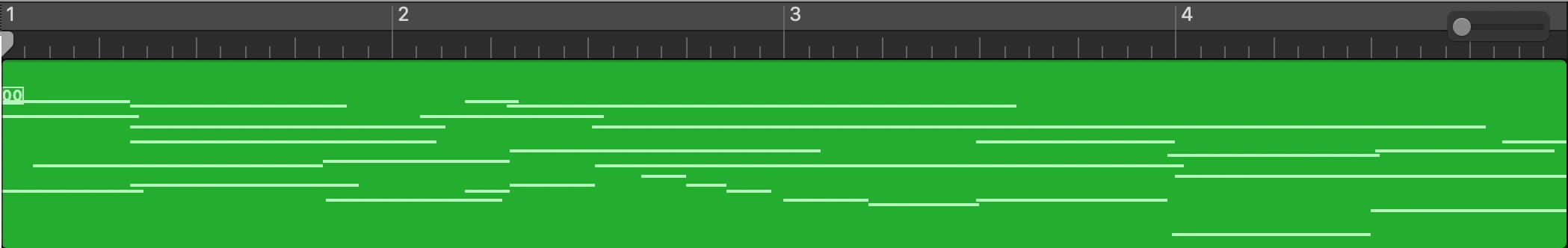}}
\caption{Three unconditionally generated piano rolls. Each piano roll in this figure is linked to the synthesized audio. To take a listen, click on one of the piano rolls. The synthesized audio can also be found at: \url{https://youtu.be/wu_kfcpzAPI}, \url{https://youtu.be/r3Rn91t3g5A}, \url{https://youtu.be/KEAtbzAScYY}.}
\label{fig:uncondtional}
\end{figure}

\begin{figure}[h]
\centering
\href{https://youtu.be/-Z5FaaligNg}{\includegraphics[width=0.99\columnwidth]{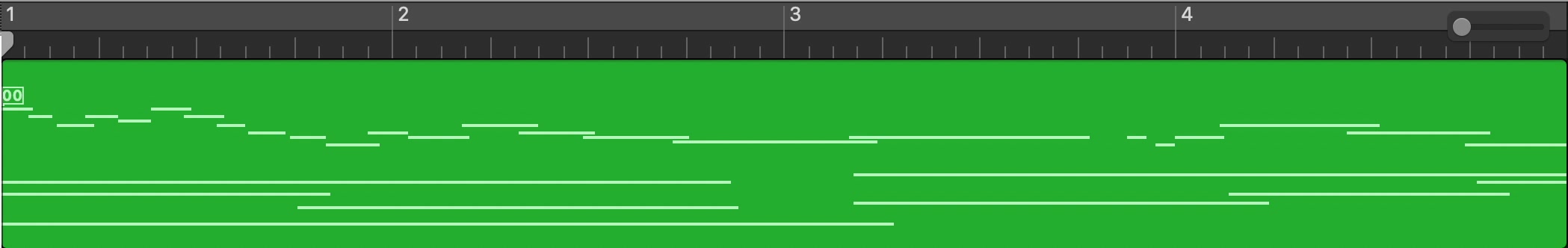}}
\href{https://youtu.be/pgnU8VLQNyQ}{\includegraphics[width=0.99\columnwidth]{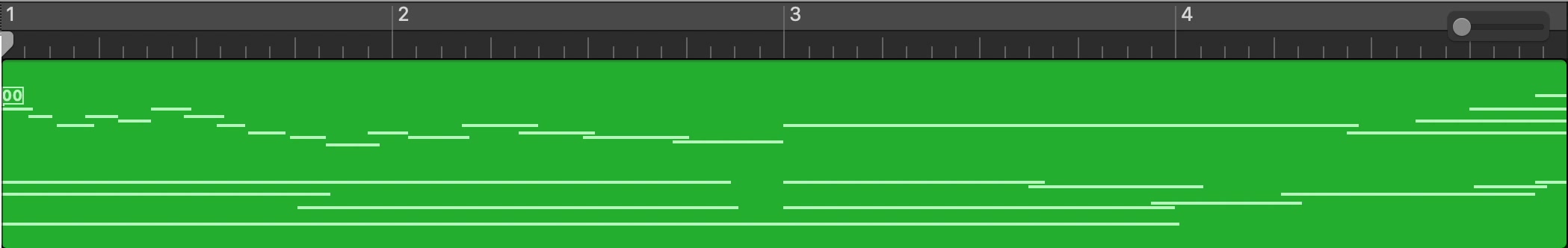}}
\caption{Top: original piano-roll segment. In the following experiments the same piano roll is used as well. Bottom: the diffusion model is prompted with the first half of the original piano roll and the second half is generated by the model. As the sampling process is stochastic, many samples can be generated using the same prompt. Here, a sample are presented. Each piano roll in this figure is linked to the synthesized audio. To take a listen, click on one of the piano rolls. The synthesized audio can also be found at: \url{https://youtu.be/-Z5FaaligNg}, \url{https://youtu.be/pgnU8VLQNyQ}.}
\label{fig:prompt_first_half}
\end{figure}

\section{Experiments}

A diffusion model can be used to generate unconditional samples following the sampling method in Algorithm~\ref{alg:sample}. Three generated piano rolls using this method are presented in Figure~\ref{fig:uncondtional}. Each piano roll in this and following figures is linked to the synthesized audio. Each of the four bars in the piano rolls is marked at the top of the pictures.

One of the interesting aspects of diffusion models is that they can be used as generative models conditioned on any part of the piano roll. For instance, it is possible to prompt the model with the first half of the piano roll and the model generates the remaining half. To achieve this, the sampling algorithm is modified by excluding the part containing the prompt from the process that adds noise to the piano roll. The algorithm begins by generating a sample piano roll from the binomial distribution. It replaces the first half of the noise sample with the prompt. The network takes this piano roll and attempts to remove the noise. Then, a smaller amount of noise is added to the piano roll, as described in the sampling algorithm. After that, the original prompt segment is placed back on the first half of the noisy piano roll, and the process repeats. Therefore, the only change to the regular sampling algorithm is that at the end of each iteration, the prompt segment is overwritten on the piano roll. At the end, the model generates a piano roll that is coherent with the prompt. The example in Figure~\ref{fig:prompt_first_half} demonstrates this approach to infilling piano rolls. The original piano roll that the model is conditioned on is not in the training set of the model. The same piano roll is used in the following experiments with conditioning the model.

\begin{figure}[h]
\centering

\href{https://youtu.be/0_xM3XFjd1w}{\includegraphics[width=0.99\columnwidth]{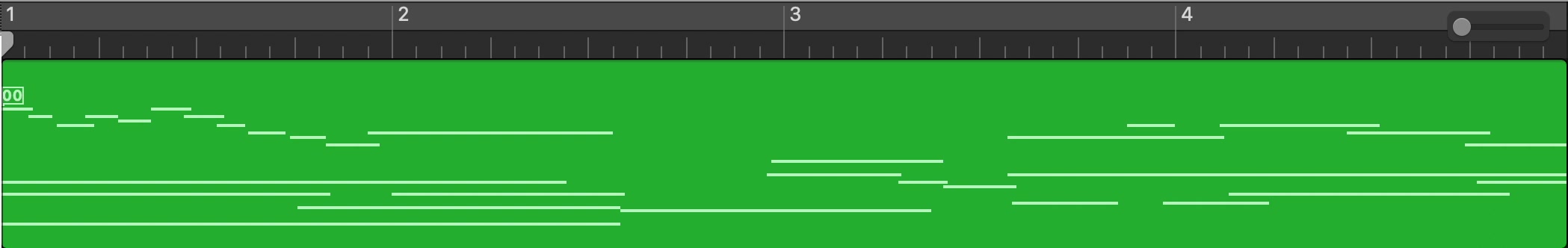}}

\caption{The model fills the middle half of the piano roll to be coherent with the first and last quarter. The piano roll in this figure is linked to the synthesized audio, which can be found at: \url{https://youtu.be/0_xM3XFjd1w}.}
\label{fig:prompt_first_last_quarter}
\end{figure}

An experiment similar to the previous one, using the conditional aspect of diffusion models, is to prompt the beginning and end of a piano roll, and then the model fills the middle segment. I thought this would be an interesting experiment to see how close would the generated style be to the original score composed by the composer. The process is similar to the previous experiment that the prompt segments are overwritten on the piano roll at the end of the sampling iterations. Figure~\ref{fig:prompt_first_last_quarter} presents a generated piano roll using this process.

\begin{figure}[h]
\centering

\href{https://youtu.be/vvpXlKrablk}{\includegraphics[width=0.99\columnwidth]{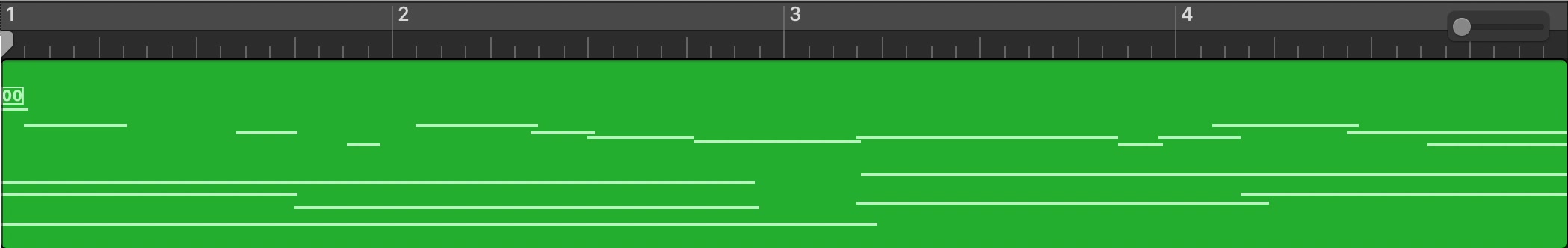}}

\href{https://youtu.be/zukJpDQOUPI}{\includegraphics[width=0.99\columnwidth]{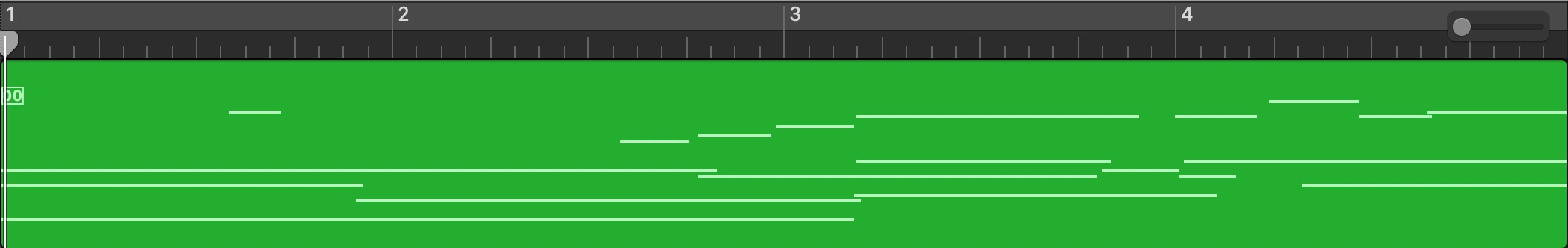}}

\href{https://youtu.be/KC4qDVuU2p0}{\includegraphics[width=0.99\columnwidth]{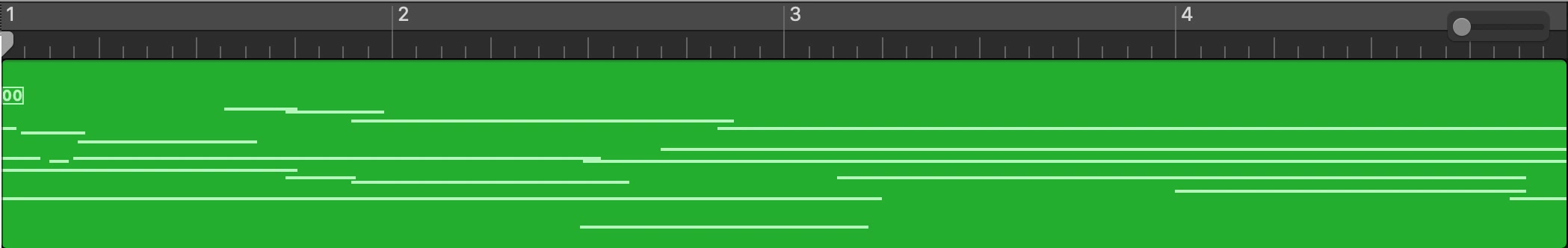}}

\caption{From top to bottom, the initial points are at $t=25,75,95$ in the sampling algorithm to generate variations of the original given piano roll. The sampling method with fewer iterations generates output that is more similar to the the given noisy piano roll. The bottom sample hardly resembles the original piano roll. The top sample sounds similar to the original piano roll. Each piano roll in this figure is linked to the synthesized audio, which can be found at: \url{https://youtu.be/vvpXlKrablk}, \url{https://youtu.be/zukJpDQOUPI}, \url{https://youtu.be/KC4qDVuU2p0}.}
\label{fig:variation}
\end{figure}

Another aspect of diffusion models is that in the latent space, which is the noisy piano roll in my experiments, pieces that sound similar are close in the latent space. Therefore, it is possible to add noise to a piano roll, for instance, the noise level of step 80 out of 100 in the forward process, and then denoise the sample using the diffusion model sampling method. The difference is that the starting point of the sampling algorithm is not a sample from the binomial distribution, but rather a piano roll with added noise at the level corresponding to the step 80 of the forward process, for instance. The output would have similarities to the original piano roll and can possibly be called a variation of the original piano roll. As the sampling process is stochastic, it is possible to generate multiple variations of a single piano roll. Using the same original piano roll as in the previous experiments, three variations of it are generated and shown in Figure~\ref{fig:variation}.

\begin{figure}[ht]
\centering
\href{https://youtu.be/unN9BBbpPOE}{\includegraphics[width=1.0\columnwidth]{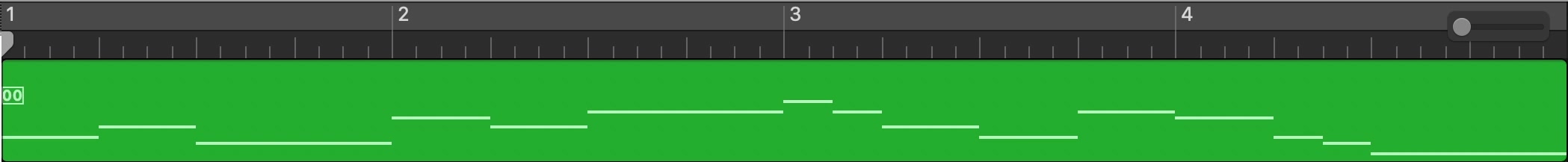}}
\href{https://youtu.be/QVDB1UfW9dc}{\includegraphics[width=1.0\columnwidth]{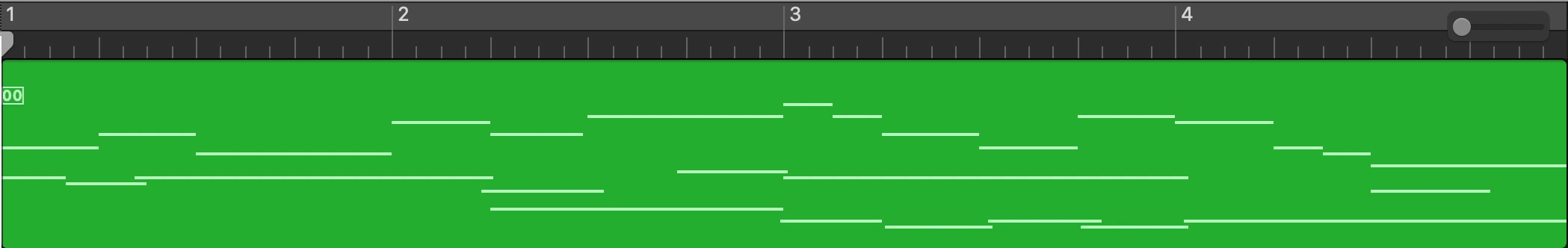}}
\caption{The trained diffusion model can generate harmony for a given melody. For this example, I improvised the melody line, which is shown in the piano roll at the top. The generated harmony, shown in the piano roll at the bottom, is created by the diffusion model and sounds idiomatic in the style of common practice period composers utilizing standard harmonic progressions. Each piano roll in this figure is linked to the synthesized audio, which can be found at: \url{https://youtu.be/unN9BBbpPOE}, \url{https://youtu.be/QVDB1UfW9dc}. \label{fig:harmonize}}
\end{figure}

It is also possible to prompt the network along the pitch axis, i.e., in each iteration of the sampling algorithm some rows are overwritten by the prompt. For example, one can provide a melody line to be harmonized by the model. Here is an example of this experiment where an improvised melody is given to the model. The melody and the combined melody and harmony piano rolls are presented in Figure~\ref{fig:harmonize}. In my subjective evaluation, the current model shows proficient harmonization and chord transitions, following common practice era principles. The generated piano rolls also display plausible modulations and cadences. For infilling and prompting tasks, the model generally performs well with no obvious dissonance or unrealistic output. However, subjective evaluation in the latter tasks is challenging due to familiarity with the original composition.

\section{Conclusions}
Binomial diffusion models can be used directly to generate piano rolls. This paper proposes an efficient forward binomial kernel for diffusion models. This kernel reduces the computational resources needed during the training process of the model. An improved sampling method for binomial diffusion models is also introduced that can more consistently produce high quality samples. Multiple generated piano rolls using the proposed method are presented to demonstrate the model can be prompted with piano-roll segments and the user can impose a melody on the piano roll which the model harmonizes. 

One drawback of diffusion models that generate piano rolls directly instead of producing vectors in the embedding space of a VAE is the increased computational complexity of the training process. For instance, training a model to generate a 32-beat long piano roll would require over 30 days on the same computer used for the experiments presented in this paper. Therefore, an interesting research direction involves enhancing the efficiency of the model and training algorithm to enable direct generation of longer piano rolls, without relying on the embedding space of an autoencoder.

\begin{acknowledgments}
I would like to express my gratitude for Miller Puckette's help in discussing the ideas that led to the work presented in this paper. Additionally, I would like to thank NVIDIA Corporation for donating the A6000 graphics cards used in this research.
\end{acknowledgments}

\bibliography{smc2023bib}

\begin{thebibliography}{10}
\providecommand{\url}[1]{#1}
\csname url@samestyle\endcsname
\providecommand{\newblock}{\relax}
\providecommand{\bibinfo}[2]{#2}
\providecommand{\BIBentrySTDinterwordspacing}{\spaceskip=0pt\relax}
\providecommand{\BIBentryALTinterwordstretchfactor}{4}
\providecommand{\BIBentryALTinterwordspacing}{\spaceskip=\fontdimen2\font plus
\BIBentryALTinterwordstretchfactor\fontdimen3\font minus
  \fontdimen4\font\relax}
\providecommand{\BIBforeignlanguage}[2]{{%
\expandafter\ifx\csname l@#1\endcsname\relax
\typeout{** WARNING: IEEEtran.bst: No hyphenation pattern has been}%
\typeout{** loaded for the language `#1'. Using the pattern for}%
\typeout{** the default language instead.}%
\else
\language=\csname l@#1\endcsname
\fi
#2}}
\providecommand{\BIBdecl}{\relax}
\BIBdecl

\bibitem{smc3}
M.~Mirza, B.~Xu, D.~Warde-Farley, S.~Ozair, A.~Courville, Y.~Bengio, I.~J.
  Goodfellow, and J.~Pouget-Abadie, ``Generative adversarial nets,''
  \emph{Advances in neural information processing systems}, vol.~27, pp.
  2672--2680, 2014.

\bibitem{smc2}
D.~P. Kingma, M.~Welling \emph{et~al.}, ``An introduction to variational
  autoencoders,'' \emph{Foundations and Trends{\textregistered} in Machine
  Learning}, vol.~12, no.~4, pp. 307--392, 2019.

\bibitem{smc4}
J.~Sohl-Dickstein, E.~Weiss, N.~Maheswaranathan, and S.~Ganguli, ``Deep
  unsupervised learning using nonequilibrium thermodynamics,'' in
  \emph{International Conference on Machine Learning}.\hskip 1em plus 0.5em
  minus 0.4em\relax PMLR, 2015, pp. 2256--2265.

\bibitem{smc5}
H.-W. Dong and Y.-H. Yang, ``Convolutional generative adversarial networks with
  binary neurons for polyphonic music generation,'' \emph{ISMIR}, 2018.

\bibitem{smc6}
A.~Roberts, J.~Engel, C.~Raffel, C.~Hawthorne, and D.~Eck, ``A hierarchical
  latent vector model for learning long-term structure in music,'' in
  \emph{International conference on machine learning}.\hskip 1em plus 0.5em
  minus 0.4em\relax PMLR, 2018, pp. 4364--4373.

\bibitem{smc7}
G.~Mittal, J.~Engel, C.~Hawthorne, and I.~Simon, ``Symbolic music generation
  with diffusion models,'' \emph{ISMIR}, 2021.

\bibitem{smc8}
R.~Rombach, A.~Blattmann, D.~Lorenz, P.~Esser, and B.~Ommer, ``High-resolution
  image synthesis with latent diffusion models,'' in \emph{Proceedings of the
  IEEE/CVF Conference on Computer Vision and Pattern Recognition}, 2022, pp.
  10\,684--10\,695.

\bibitem{smc9}
J.~Ho, A.~Jain, and P.~Abbeel, ``Denoising diffusion probabilistic models,''
  \emph{Advances in Neural Information Processing Systems}, vol.~33, pp.
  6840--6851, 2020.

\bibitem{smc10}
A.~Bansal, E.~Borgnia, H.-M. Chu, J.~S. Li, H.~Kazemi, F.~Huang, M.~Goldblum,
  J.~Geiping, and T.~Goldstein, ``Cold diffusion: Inverting arbitrary image
  transforms without noise,'' \emph{arXiv preprint arXiv:2208.09392}, 2022.

\bibitem{smc11}
O.~Ronneberger, P.~Fischer, and T.~Brox, ``U-net: Convolutional networks for
  biomedical image segmentation,'' in \emph{International Conference on Medical
  image computing and computer-assisted intervention}.\hskip 1em plus 0.5em
  minus 0.4em\relax Springer, 2015, pp. 234--241.

\bibitem{smc12}
D.~Erhan, Y.~Bengio, A.~Courville, and P.~Vincent, ``Visualizing higher-layer
  features of a deep network,'' \emph{University of Montreal}, vol. 1341,
  no.~3, p.~1, 2009.

\end{thebibliography}

\end{document}